\documentclass[doublecol]{epl2}

\usepackage{graphicx}

\newcommand{\lbbrack}{\Bigg[\!\!\!\!\Bigg[}
\newcommand{\rbbrack}{\Bigg]\!\!\!\!\Bigg]}

\title{New Quantum Limits in Plasmonic Devices}

\author{M. Marklund\inst{1} \and G. Brodin\inst{1} \and L. Stenflo\inst{1} \and C.S. Liu\inst{2}} 
\shortauthor{M. Marklund \etal}

\institute{
\inst{1} Department of Physics, Ume{\aa} University, SE--901 87 Ume{\aa}, Sweden \\
\inst{2} Department of Physics, University of Maryland, College Park, Maryland 20742, USA
}


\abstract{
Surface plasmon polaritons (SPPs) 
have recently been recognized as an important future technique for 
microelectronics. Such SPPs have been studied using classical theory. However, 
current state-of-the-art experiments are rapidly approaching nanoscales, and quantum effects 
can then become important. 
Here we study the properties of quantum SPPs at the interface between an
electron quantum plasma and a dielectric material. It is shown that
the effect of quantum broadening of the transition layer is most important.
In particular, the damping of SPPs does not vanish even in the
absence of collisional dissipation, thus posing a fundamental size limit for plasmonic
devices. Consequences and applications of our results are pointed out.
}

\begin{document}

\maketitle

%\paragraph*{Introduction}

The excitation and propagation of surface modes at plasma interfaces have
long been important in space physics, magnetic confinement fusion, and laboratory plasma physics 
\cite{gradov-stenflo,yu,stenflo,aliev-etal}. 
Moreover, studies of electron
oscillation excitations (surface plasmons, surface plasmon polaritons (SPPs) and magnetoplasmons) \cite{feibelman73,feibelman74,tsuei-etal} in nano-structured systems \cite{holland-etal,sukhodub-etal,kneipp,chang-etal} 
have recently attracted much interest.
It has been found that at condensed matter interfaces,
such plasmon excitations could be of crucial importance in future electronic components 
\cite{lezec-etal,barnes-etal,maier-atwater,maier}, the latter referred to as plasmonic
devices \cite{ozbay,pitarke-etal}.

%In the 1960's, Pines studied the excitation spectrum of quantum plasmas \cite
%{Pines}, with a high density and a low temperature as compared to normal
%plasmas. In such systems, the finite width of the electron wave function
%makes quantum tunnelling effects crucial, leading to an altered dispersion
%relation. The field of quantum plasmas has since then undergone a rapid
%development \cite{manfredi}. 

%Recently, a number or studies of the quantum
%properties of plasmas have been published (see, \textit{e.g.}, Ref.\ %
%\onlinecite{kremp-etal} and references therein), including electron spin effects 
%\cite{marklund-brodin}.

Furthermore, the field of quantum plasmas has recently developed rapidly \cite
{manfredi}. This field started already in the 1960's, when Pines
studied the excitation spectrum of quantum plasmas \cite{Pines}, with
a high density and a low temperature as compared to normal plasmas.
In such systems, the finite width of the electron wave function makes
quantum tunnelling effects crucial, leading to an altered dispersion
relation. Since then, a number of theoretical studies of quantum statistical
properties of plasmas have been published (see, \textit{e.g.}, Ref.\ \cite{kremp-etal}
and references therein). For example, Bezzerides \& DuBois presented a
kinetic theory for the quantum electrodynamical properties of nonthermal
plasmas \cite{bezzerides-dubois}, while Hakim \& Heyvaerts used a covariant
Wigner function approach for relativistic quantum plasmas \cite
{hakim-heyvaerts}. It has also been shown that, under certain conditions,
plasmas can display remarkable properties due to the quantum properties of
the constituents. Thus, there are  quantum multistream instabilities 
\cite{haas-etal,anderson-etal}, quantum modified Zakharov dynamics \cite{garcia-etal,marklund}  
together with soliton formation and nonlinear quantum interactions \cite
{shukla-eliasson,shukla-eliasson2}, spin effects on the plasma dispersion 
\cite{marklund-brodin,brodin-marklund}, quantum plasma turbulence \cite
{shaikh-shukla}, ferromagnetic plasma behaviour and Jeans-like
instabilities due to quantum effects \cite{brodin-marklund2}.

Many of the current studies involving quantum plasmas are motivated by the
rapid experimental progress and development of new materials, \textit{e.g.},
nanostructured materials \cite{craighead} and quantum wells \cite
{manfredi-hervieux}, and the laboratory realization of ultracold plasmas 
\cite{robinson-etal} and experimental demonstration of quantum plasma
oscillations in Rydberg systems \cite{fletcher-etal}. Quantum dispersive
effects can also be important for diagnostics of inertial fusion
plasmas \cite{glenzer-etal}. In parallel, the field of plasmonics and its
use of surface waves, such as surface plasmon polaritons
(SPPs), has emerged as a new route to electronic devices \cite
{atwater}.

%Thus, the
%merging of the fields of plasmonics, with its high application potential,
%and the rapidly growing field of quantum plasmas at the interface of small
%scale electronic devices will yield new and potentially important results.

In this Letter, methods from quantum plasma physics are used for analyzing
SPPs in nanoscale systems. In particular, we determine the dispersion
relation for quantum SPPs on a conductor-dielectric
interface. It is shown that wave function dispersion
introduces an intrinsic damping, even in the absence of collisions. Such
damping is due to the irreversible propagation of resonant plasmons 
towards lower density regions. This is of importance for the short length scales in
forthcoming electronic components based on SPPs in
waveguide slots. An expression for the damping length, limiting the size of
such devices, is presented. For wavelengths of the order of tens of
nanometers, the propagation distance for the SPPs is only a fraction of the
wavelength.

%%%%%%
%We are interested in electron plasma oscillations on material
%surfaces/interfaces. Thus, we assume that the conducting medium is
%semi-infinite in the positive $x$-direction, while the dielectric is
%semi-infinite in the negative $x$-direction. The stationary background
%electron density $n_0$ and the permittivity $\epsilon_{\mathrm{d}}$ of the
%dielectric is then a function of $x$ (see Fig.\ 1). 

%%%%%%%%% INSERTED

%\paragraph*{Governing equations.}

The governing equations for the electrostatic electron dynamics are \cite{manfredi} 
%\begin{equation}  \label{eq:cont-full}
the continuity equation $\partial_t n + \nabla\cdot(n\mathbf{v}) = 0$, 
%\end{equation}
the momentum conservation equation
\begin{equation}  \label{eq:mom-full}
mn(\partial_t + \mathbf{v}\cdot\nabla)\mathbf{v} = -en\mathbf{E} -\nabla p -
mn\nu\mathbf{v}+ \frac{\hbar^2n}{2m}\nabla\left( \frac{\nabla^2\sqrt{n}}{%
\sqrt{n}} \right),
\end{equation}
and Poisson's equation
%\begin{equation}  \label{eq:poisson-full}
 $\nabla\cdot(\epsilon_0\mathbf{E} + \mathbf{P})= n_i - n$, 
%\end{equation}
where $n$ is the electron density, $n_i$ is the ion density (here we treat
the lattice constituents in terms of the ion density, in order to
demonstrate the analogy to classical plasmas and surface plasmons), $m$ is
the electron mass, $\mathbf{v}$ is the electron velocity, $e$ is the
magnitude of the electron charge, $\mathbf{E}$ denotes the electric field
strength, $p$ is the electron pressure, $\hbar$ is Planck's constant, $%
\epsilon_0$ is the vacuum dielectric constant, $\nu$ is the collisional
frequency between the electrons and the ions, $\mathbf{P} =
\epsilon_0(\epsilon_{\mathrm{d}} - 1)\mathbf{E}$ is the polarization due to
bound charges in the system, and $\epsilon_{\mathrm{d}}$ is the
permittivity. The last term in Eq.\ (1), the gradient
of the Bohm--de Broglie potential, corresponds to
the effect of wave function dispersion. 

We consider the ions as stationary, and write the perturbed electric
field according to $\delta\mathbf{E} = -\nabla\delta\phi$. Moreover, we
assume that the conducting medium is semi-infinite in the positive $x$%
-direction, so that the stationary background electron density $n_0$ is a
function of $x$ (see Fig.\ 1). 
%The electron density fluctuations $\delta n$
%then satisfy the linearized continuity equation 
%\begin{equation}  \label{eq:cont}
%\partial_t\delta n + \nabla\cdot(n_0\delta\mathbf{v}) = 0 ,
%\end{equation}
%where the electron velocity perturbation $\delta\mathbf{v}$ satisfies the
%momentum conservation equation 
%\begin{equation}  \label{eq:mom}
%mn_0\partial_t\delta\mathbf{v} = en_0\nabla\delta\phi - m\nabla(c_s^2\delta
%n) - mn_0\nu\delta\mathbf{v} + mn_0\nabla \delta U_B,
%\end{equation}
%and the electrostatic potential is determined by 
%\begin{equation}  \label{eq:poisson}
%\nabla\cdot\left[\epsilon_{\mathrm{d}}\nabla\delta\phi \right] = \frac{%
%e\delta n}{\epsilon_0} .
%\end{equation}
%Here $c_s = [(dp/dn)|_{n_0}/m]^{1/2}$ is the electron sound speed. We will
%from here on, when necessary, consider a pressure $p =
%(4\pi^2\hbar^2/5m)(3/8\pi)^{2/3}n^{5/3}$ of a non-relativistic degenerate
%electron gas, so that the sound speed becomes $c_s = (2\pi\hbar/\sqrt{3}%
%\,m)(3n_0/8\pi)^{1/3}$. The perturbed Bohm--de Broglie potential is given by 
%$\delta U_B = (\hbar^2/4m^2n_0)\nabla\cdot(n_0\nabla)\delta n/n_0 $.
%Following the standard procedure, \textit{i.e.}, assuming the form $%
%g(x)\exp(iky - i\omega t)$ of all perturbed quantities \cite{kaw-mcbride} we
%obtain Eq.\ (\ref{eq:final}). 
%%%%%%%%%%%%%%%
Next, we linearize the governing equations. 
%and obtain the
%equation determining the SPP dispersion relation. 
In the cold classical case we obtain
the SPP dispersion relation
$\nabla\cdot[\epsilon(x)\nabla\delta\phi] = 0$, where $\epsilon(x) =
\epsilon_{\mathrm{d}}(x) - \omega_{\mathrm{p}}^2(x)/\omega(\omega + i\nu) $
is the total dielectric function for a cold classical plasma with collision
frequency $\nu$, and $\omega_{\mathrm{p}}(x) = [e^2n_0(x)/\epsilon_0m]^{1/2}$
is the electron plasma frequency. We note that the collisional frequency is
a function of the ion density. In the general case, we obtain 
\begin{equation}  \label{eq:final}
\partial_a\big[ \hat{\epsilon}^{ab}\partial_b\delta\phi \big] = 0
\end{equation}
where we use the index notation $\nabla \rightarrow \partial_a$ etc., as
well as Einstein's summation convention and let all perturbed quantities be
of the form $g(x)\exp(iky - i\omega t)$ \cite{kaw-mcbride}. 
Here the dielectric tensor operator
is given by 
\begin{equation}
\hat{\epsilon}^{ab} = \delta^{ab}\epsilon + \lbbrack \frac{\partial^ac_s^2}{\omega^2} - 
\frac{\omega_{\mathrm{p}}^2\partial^a}{4m^2}\left\{ \frac{\hbar^2\partial_c}{\omega_{\mathrm{p}}^2}%
\left[\frac{\omega_{\mathrm{p}}^2\partial^c}{\omega^2}\left(\frac{1}{\omega_{%
\mathrm{p}}^2}\right) \right] \right\} \rbbrack\partial^b\epsilon_{%
\mathrm{d}} ,  \label{eq:operator}
\end{equation}
where $c_s = [(dp/dn)|_{n_0}/m]^{1/2}$ is the sound speed. We will from here on consider a pressure $p =
(4\pi^2\hbar^2/5m)(3/8\pi)^{2/3}n^{5/3}$ of a non-relativistic degenerate
electron gas, so that the sound speed becomes $c_s = (2\pi\hbar/\sqrt{3}%
\,m)(3n_0/8\pi)^{1/3}$. 
%We note
%that all non-diagonal terms are due to quantum effects.
%, \textit{i.e.}, the
%Fermi pressure and the Bohm--de Broglie potential, and that a finite temperature 
%would introduce a further diagonal term in the dielectric tensor. However, since the inclusion of such a
%classical effect is trivial, we will here focus on the Fermi pressure.

%%%%%

%%%%%% FIGURE DIELECTRIC%%%%%%
\begin{figure}[tbp]
\includegraphics[width=.8\columnwidth]{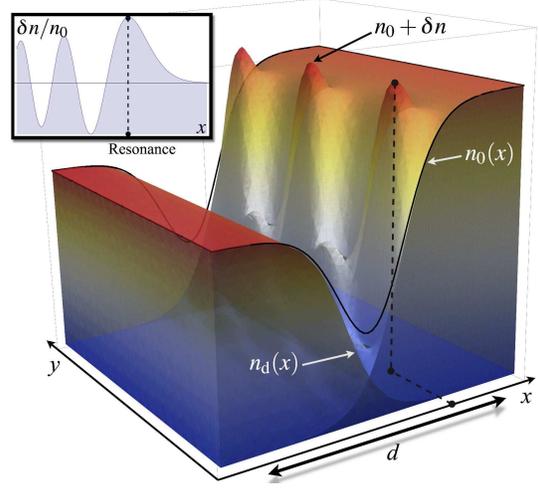}
\caption{The structure of the electron density profile for the SPP
excitation. The width $d$ of the transition layer is determined by quantum
effects. A dielectric material, with electron density 
$n_\mathrm{d}(x)$, is on the left, while the metal with the corresponding plasma
oscillations is on the right. The small panel in the upper left corner shows the
Airy-like structure of the density perturbations $\protect\delta n/n_0$.}
\end{figure}
%%%%%%%%%%%%%%%%%

We consider a plasma which can be divided into three regions (see Fig.\ 1).
For $x<0$, the unperturbed plasma density is zero. For $0\leq x\leq d$, we
have a transition layer where the plasma density is a monotonously
increasing function of $x$. Lastly, the plasma density is constant for $x>d$%
. Furthermore, for the surface waves to be only weakly damped, the condition 
$kd\ll 1$ must be fulfilled. The physics inside the transition layer is then
essentially one-dimensional, since $k\ll \partial /\partial x$ applies here.

%%%%%% INSERTED

%\paragraph{The transition layer.}

Integrating Eq.\ (\ref{eq:final}) across the transition layer, we
obtain 
\begin{equation}
\big[\hat{\epsilon}^{xx}\partial _{x}\delta \phi +ik\hat{\epsilon}%
^{xy}\delta \phi \big]_{0}^{d}=\int_{0}^{d}\left[ k^{2}\hat{\epsilon}%
^{yy}\delta \phi -ik\hat{\epsilon}^{yx}\partial _{x}\delta \phi \right] \,dx.
\label{Eq:BC1}
\end{equation}
Moreover, $\delta \phi (d)$ and $\delta \phi (0)$ are related by 
\begin{equation}
\delta \phi (d)-\delta \phi (0)=\int_{0}^{d}\frac{\partial \delta \phi }{%
\partial x}\,dx= -\frac{ie}{\omega}\int_{0}^{d}n_{0}(x)v_{x}(x)\,dx
\label{Eq:BC2}
\end{equation}
The integrals in Eqs. (\ref{Eq:BC1}) and (\ref{Eq:BC2}) are correction terms
proportional to $kd$. Similarly the off-diagonal quantum terms in the left
hand side of (\ref{Eq:BC1}) are also proportional to one of two small
parameters, namely $k^{4}\hbar ^{2}/m^{2}\omega_{\mathrm{p}}^{2}$ (from the
Bohm--de Broglie potential) or $k^{2}c_{s}^{2}/\omega_{\mathrm{p}}^{2}$
(from the Fermi pressure). If we drop both these quantum terms and let $d\rightarrow 0$,
we obtain the wellknown electrostatic
dispersion relation $\omega =\omega_{\mathrm{p}%
}^{(0)}(1+\epsilon_{\mathrm{d}}^{(0)})^{-1/2}$, where $\epsilon_{\mathrm{d}%
}^{(0)} = \epsilon_\mathrm{d}(x < 0)$ is the constant dielectric
permittivity for $x<0$, and $\omega_{\mathrm{p}}^{(0)}$ is the plasma
freqency at $x=d$. Within this approximation, the electrostatic surface wave
has a zero damping as well as a zero group velocity. The quantum terms
modify this result, since $d$ now remains finite. 
%Similar terms exist also in a classical
%context, provided the ion density profile has a finite width. Within our
%zero temperature approximation, however, a sharp
%unperturbed ion density implies an identical electron profile, unless the
%quantum terms smearing out the electrons are included. 
To adress this case properly, we must determine $n_{0}(x)v_{x}(x)$ inside the
transition layer. Using the approximation $k\ll \partial /\partial x$ inside
this region we find 
\begin{equation}
\hat{\epsilon}^{xx}[n_{0}(x)v_{x}(x)]=-\frac{i\omega k}{e}\hat{\epsilon}%
^{xx}(0)\delta \phi (0) .  \label{Eq:General-layer}
\end{equation}

We can here
treat $\delta \phi (x)$ as a constant within the layer, in which case the right hand
side of (\ref{Eq:BC1}) reduces to $\delta \phi (0)\int_{0}^{d}k^{2}\epsilon
(x)\,dx$. Moreover, when the inhomogeneous derivatives on the left hand
side of (\ref{Eq:BC1}) vanish, the remaining quantum terms become
proportional to $\nabla ^{2}\delta \phi $. However, for $k^{2}c_{s}^{2}/%
\omega _{\mathrm{p}}^{2}\ll 1$ and $k^{4}\hbar ^{2}/m^{2}\omega _{\mathrm{p}%
}^{2}\ll 1$, the homogeneous
regions are classical to first order, \textit{i.e.}, $\nabla ^{2}\delta \phi =0$. 
%We note that the homogeneous quantum terms would
%have been small as compared to quantum effects inside the transition layer, even
%if the terms had not vanished to all orders. The general dispersion relation
%(\ref{Eq:Final-result}) follows from the above discussion. 
%
%%%%%%%%%%%%%
%
%
%Integrating Eq. (\ref{eq:operator}) twice across the transition layer to
%find the boundary conditions, and considering exponentially decaying
%solutions away from the boundary, 
The dispersion relation for SPPs then becomes 
\begin{equation}
\omega =\frac{\omega _{\mathrm{p}}^{(0)}}{(1+\epsilon _{\mathrm{d}%
}^{(0)})^{1/2}}\left[ 1+\frac{k}{2}\int_{0}^{d}[\epsilon (x)+q(x)]\,dx\right]
\label{Eq:Final-result}
\end{equation}
where $q(x)$ is the solution to Eq.\ (\ref{Eq:General-layer}) for the
plasmon field inside the transition layer, defined by $%
q(x)=ien_{0}(x)v_{x}(x)/\omega k\delta \phi (0)$.

The integral contribution in (\ref{Eq:Final-result}) depends on the equilibrium profile. Before
turning our attention to the full quantum equilibrium profile we consider the following. 

i) \textit{Classical case.} Here a finite width of the electron distribution in the
transitions layer is determined by a finite width of the ion profile.
From (\ref{Eq:General-layer}) we
obtain $q(x)=\epsilon (0)/\epsilon (x)$. The singularity in this equation
can be dealt with using the Landau prescription, which results in an
imaginary contribution to the integral, $\mathrm{Im}\{\int_{0}^{d}qdx\}=i\pi
\epsilon (0)(\partial \epsilon /\partial x)_{\epsilon =0}^{-1}$ in Eq.\ (7), where the
subscript indicates that the derivative should be evaluated at the resonant
surface where $\epsilon =0$. The imaginary part representing damping of the
SPP due to energy transfer to the resonant plasmons in the
transition layer.

ii) \textit{Fermi pressure effects.} Assuming that the width is determined by
the ion-density profile we drop the Bohm--de Broglie potential. In this
case, (\ref{Eq:General-layer}) has the solution $q(x)= \mathrm{Gi}%
(xd\epsilon /dx)+i\mathrm{Ai}(xd\epsilon /dx)$, where we for
simplicity assume that $\epsilon (x)$ is a linear function. Here $\mathrm{Gi}$ and
$\mathrm{Ai}$ denote solutions to the inhomogeneous and homogeneous Airy equation \cite{Abromowitz-Stegun}, 
respectively.The proper causual solution
has been found by taking the nondivergent solution to Eq.\ (\ref{Eq:General-layer}) that
represents resonant plasmons propagating towards lower densities. Due to negligible
reflection of the plasmons at the dielectric interface, the energy is
irreversibly lost, even in the absence of dissipation. As a consequence,
 the damping of the surface wave%, determined by $\mathrm{Im}% [\int_{0}^{d}qdx]$, 
 coincides between cases i and ii.
%provided the expression for $\epsilon (x)$ in both cases is the same close
%to the resonance $\epsilon (x)=0$. %%%%%%%%
%The proper causual solution has been found by taking the
%nondivergent solution to Eq.\ (\ref{Eq:General-layer})that represents a Langmuir wave propagating towards
%the lower density side. Interestingly, although the solution for the
%velocity field differs significantly between cases \textit{i} and \textit{ii}, the damping of
%the surface wave, determined by $\mathrm{Im}[\int_{0}^{d}n_{0}(x)v_{x}dx]$,
%coincides in the cases \textit{i} and \textit{ii}, provided the expression for $\epsilon (x)$
%is the same close to the resonance $\epsilon (x)=0$.

%%%%%%% INSERTED

%\paragraph{Equilibrium solution.}

Next, we investigate
the equibrium density profile in the transition layer, determined by 
\begin{equation}
\frac{d}{d\bar{x}}\left\{ \epsilon_{\mathrm{d}}\frac{d}{d\bar{x}}\left[ \eta%
\bar{n}^{2/3} - \frac{1}{\sqrt{\bar{n}}}\frac{d^2\sqrt{\bar{n}}}{d\bar{x}^2} %
\right] \right\} = \bar{n} - H(-x) ,  \label{Eq:Equilibrium}
\end{equation}
where we have introduced the normalized distance $\bar{x}=x/(\hbar /\sqrt{2}%
\,m\omega_{\mathrm{p}}^{(0)})^{1/2}$, the normalized density $\bar{n}%
=n_0(x)/n_{0}^{(0)}$, taken the normalized ion-density as a step function $%
H(-\bar{x})$, and introduced the dimensionless parameter $\eta \equiv
4\pi^2(\hbar/\sqrt{2}\,m\omega_{\mathrm{p}}^{(0)})(3n_0^{(0)}/8\pi)^{2/3}$.
%The first term in the left hand side is due to the Bohm--de Broglie
%potential, whereas the second term is due to the Fermi pressure. Since both $%
%\sqrt{\bar{n}}$ and $\bar{n}^{2/3}$ varies from unity to zero across the
%transition layer, clearly 
The relative importance of the the Fermi pressure
and the Bohm--de Broglie potential is determined by the parameter $\eta$. We
first consider the case $\eta \ll 1$, \textit{i.e.}, the Bohm--de Broglie
potential dominates over the Fermi pressure. Noting that $\bar{n}-H(-\bar{x})
$ is of order unity after passing the ion density step, the normalized width 
$\bar{d}$ must be of order unity, \textit{i.e.}, in dimensional units we
have $d \sim (\hbar /m\omega_{\mathrm{p}}^{(0)})^{1/2}$. On the other hand,
assuming a dominating Fermi pressure, \textit{i.e.}, $\eta \gg 1$, the width
becomes $d\sim ( \eta \hbar /m\omega_{\mathrm{p}}^{(0)})^{1/2}$. For a solid
state plasma, which is the case of most interest, $\eta $ is of order unity,
and thus we get $d$ $\sim (\hbar /m\omega_{\mathrm{p}}^{(0)})$ independently
of which of the two estimates we use. Assuming $\eta \sim 1$, 
we can deduce a number of properties of the resonant
plasmon field inside the transition layer using the solutions to (\ref{Eq:General-layer}).

%%%%%%%%%%%%%

With $n_{0}(x)$ determined by (\ref{Eq:Equilibrium}), we can now treat the
full quantum case. We start with the simple observation that $q(x)$ is
independent of $k$, as terms involving $k$ can be neglected in (\ref
{Eq:General-layer}) due to $k\ll \partial /\partial x$. Solving Eq.\ (\ref
{Eq:General-layer}) to find a full solution for $q(x)$ is a nontrivial task,
requiring a solution to the equilibrium equation (\ref{Eq:Equilibrium}),
that has to be found numerically. However, a qualitative understanding can
be obtained by Taylor expanding Eq.\ (\ref{Eq:General-layer}) close to the
classical plasmon resonance $\epsilon (x)=0$. We obtain an Airy like
behavior of the plasma oscillations along the $x$-direction in the
transition layer, depicted in the upper left panel of Fig.\ 1. The full
quantum case gives two major changes to the classical and Fermi pressure
cases. Firstly, the inhomogeneities in the transition layer
shifts the plasmon resonance somewhat towards higher densities. Secondly,
the typical plasmon wavelength in the $x$-direction becomes comparable to
the width of the transition layer. Using this as a basis for an estimate,
together with the condition that no plasmons propagate towards
higher densities in the transition layer, we find that $\int_{0}^{d}q(x)\,dx%
%\approx (\pi \hbar /m\omega )^{1/2}\int_{-1}^{1}[\mathrm{Gi}(x)+i\mathrm{Ai}(x)]dx
\approx (\pi \hbar /m\omega )^{1/2}\left( 0.6+2i\right) $. 
%While seemingly this is an accurate expression, it should
%be stressed that the uncertainty in this formula is significant, and
The uncertainty in this expression is directly linked to the uncertainty in
the unperturbed density profile and the expression for the transition layer
width $d$. Making a similar estimate of the term $\int_{0}^{{d}}\epsilon
(x)\,dx$ in (\ref{Eq:Final-result}), we use a linear profile for $\epsilon
(x)$, in which case we obtain $\int_{0}^{{d}}\epsilon (x)\,dx=(\pi \hbar
/m\omega )^{1/2}[\epsilon (x=d)-\epsilon (x=0)]\approx 0$, where the last
equality follows from the dispersion relation (\ref{Eq:Final-result}). A more accurate
density profile gives a finite contribution. However, the plasmon coupling term $%
\propto \int_{0}^{d}q(x)\,dx$ will still dominate. Thus, we can write our
final dispersion relation for the SPPs as 
\begin{equation}
\omega \approx \frac{\omega _{\mathrm{p}}^{(0)}}{(1+\epsilon _{\mathrm{d}%
}^{(0)})^{1/2}}\left[ 1+(0.6+2i)\left( \frac{\hbar k^{2}}{m\omega _{\mathrm{p%
}}^{(0)}}\right) ^{1/2}\right] .  \label{Eq:Final-Really}
\end{equation}
The group velocity and the damping rate of the electrostatic SPPs are $%
V_{g}=\partial \omega /\partial k=0.6V_{q}$ and $\mathrm{Im}\,\omega =2kV_{q}
$, respectively, where we have introduced the characteristic velocity $%
V_{q}=[{\hbar \omega _{\mathrm{p}}^{(0)}}/{m(1+\epsilon _{\mathrm{d}}^{(0)})}%
]^{1/2}$. In Fig.\ 2 we display the real part of the dispersion relation (%
\ref{Eq:Final-Really}). Apart from the shortest wavelengths, the dispersive properties are dominated by electromagnetic effects. In the short wavelength limit the group velocity of the classical dispersion relation approaches zero, whereas the quantum corrected value $V_g$ approaches a small but nonzero constant value $V_q$.  Let us point out that there might be classical effects not included in our model that can modify the given picture, e.g. thermal effects which  can induce a  $k$-dependence of $\epsilon$. We note, however, that within a fluid model, the density perturbation of the surface wave approaches zero, which leads to negligible modifications of our model. Within a kinetic picture, on the other hand, thermal effects induce classical modifications with a $k$-dependence of $\epsilon$, which might lead to modifications of the dispersive properties also in the short wavelength limit.

%%%%% FIG %%%%%
\begin{figure}[tbp]
\includegraphics[width=.8\columnwidth]{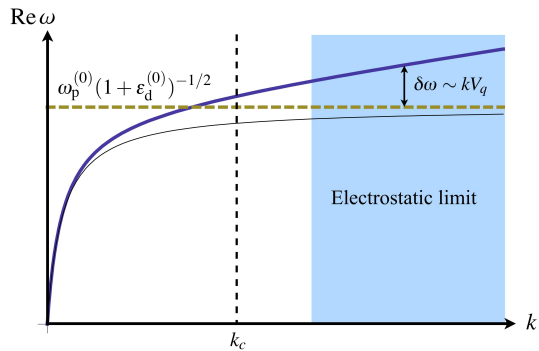}
\caption{The classical dispersion relation (thin curve), and
the quantum dispersion relation (thick curve) for the SPPs. The dashed horizontal line gives the classical resonance frequency. 
For wavenumbers larger than $k_c \equiv (\protect\epsilon_\mathrm{d}^{(0)}\protect\omega_\mathrm{p}%
^{(0)})^{5/6}(m/\hbar c^4)^{1/6} $, the quantum contribution to the group velocity is larger than the
classical electromagnetic contribution. We note that quantum damping, as
represented by $\mathrm{Im}\,\protect\omega$ in (\ref{Eq:Final-Really}), can
be important for wavenumber lower than $k_c$. }
\end{figure}
%%%%%%%%%%%%

%\paragraph*{Discussion and conclusion}

However, the main importance of the dispersion relation (9) concerns the dissipative (i.e.\ imaginary) part, which can have significant consequences for the rapidly emerging
fields of plasmonics \cite{maier-atwater,ozbay,pitarke-etal}, where plasmon
wave propagation along metallic-dielectric interfaces
is studied as a means to pave the way for even smaller and faster electronc
circuits \cite{maier,maier-atwater}. Naturally, energy losses must be
minimized if this undertaking is to be successful. Currently the propagation
distances of the SPPs range from a few 100 nanometers up to tens of microns 
\cite{maier-atwater}. The way to improve performance has so far been to
consider wavguide slots \cite{maier-atwater,ozbay,pitarke-etal}, where most
of the energy is distributed in a dielectric, limiting the collisional
losses in the metal surface. By cooling the system collisional losses can be
further minimized \cite{raether}. 
%However, our result Eq.\ (\ref{Eq:Final-Really}) poses a \emph{fundamental limit} when trying to build
%continuously smaller circuits. 
However, when reducing the size %of such devices
towards the nanoscale, the wavelength of the SPP is also decreased. Assuming
that collisional effects are minimized when approaching such a regime, our
treatment of quantum losses becomes important. For such nano-sized system, a
damping of the order $\mathrm{Im}(\omega )/|\omega |\sim k(\hbar /m\omega_{%
\mathrm{p}})^{1/2} $, as given by Eq.\ (\ref{Eq:Final-Really}), becomes
crucial, and sets a fundamental limiting factor for how small systems that
can be designed. 
%Above, we have seen that the effect of the quantum correction
%to the electron motion gives rise to a damping, due to the implied finite
%transition layer width. We next look at the effect of the quantum broadening
%in small-scale plasmonic devices. 
The damping length of the SPP due to
quantum effects is given by $\delta_{\mathrm{SP}} = V_{g,\mathrm{em}}/\mathrm{%
Im}\,\omega$, where $V_{g,\mathrm{em}} \approx (\epsilon_{\mathrm{d}%
}^{(0)}/c)^2(\omega_{\mathrm{p}}^{(0)}/k)^3$ is the group velocity including
electromagnetic effects \cite{aliev-etal} in the short wavelength region,
while $\mathrm{Im}\,\omega$ is given by (\ref{Eq:Final-Really}). It should
be stressed that the electromagnetic contribution to the group velocity
dominates over the quantum induced contribution $V_q$ for wavelengths $%
\lambda \geq 30\,\mathrm{nm}$. Assuming that the dielectric consists of $%
\mathrm{SiO}_2$, we have \cite{handbook} $\epsilon_{\mathrm{d}}^{(0)} \sim 3
- 5$, and with the plasma frequency of the metal as $\omega_{\mathrm{p}%
}^{(0)} \approx 4\times10^{15}\,\mathrm{s}^{-1}$, we obtain 
\begin{equation}  \label{eq:damping}
\delta_{\mathrm{SP}} \approx \left(\frac{\lambda}{100\,\mathrm{nm}}
\right)^4\,\mu\mathrm{m} ,
\end{equation}
where $\lambda = 2\pi/k$ is the wavelength. Thus, due to the strong
wavelength dependence in (\ref{eq:damping}), $\mu\mathrm{m}$-waves can
propagate without significant quantum damping, while decreasing the scale
much below the $\mu\mathrm{m}$ regime will affect the effective propagation
distance. For example, for $\lambda \sim 30\,\mathrm{nm}$ the damping length 
$\delta_{\mathrm{SP}} \sim 10\,\mathrm{nm}$. Although different geometries may
affect the possibilities to design smaller devices \cite{maier-atwater}, our
result (\ref{eq:damping}) is robust, and its consequences must therefore be
considered in the design of plasmonic devices. 
%Spintronic aspects is not expected to
%modify the present result significantly \cite{chau-etal}.

%In this Letter, we have determined the dispersion relation of quantum SPPs
%at a dielectric interface, using a general framework for analyzing quantum
%plasma perturbations on inhomogeneous backgrounds. In particular, we have
%shown that the effect of electron wave function dispersion is to introduce a
%damping of the surface modes. Energy is irreversibly lost due to bulk
%plasmon propagation towards lower densities in the conductor. Such a
%non-dissipative damping sets a fundamental limit to the size of forthcoming
%plasmonic devices, due to the strong dependence of the propagation length on
%the wavelength of the SPPs. The result will therefore be of importance for
%the design and construction of future plasmonic components and circuits. 
%%Generalizations to
%%electromagnetic perturbations are straightforward, and introduce no
%%important modifications to the present results of quantum induced damping. 

%%%%%%%%%%%%%%%%%%%%%%%%%%%%%%%%%%%%%%%%%%%%%%%%%%%%%%%%%%%%%%%%%%%%%%%%%%%%%%%%%%%%%%%%%%%%%%

%\acknowledgments This research was supported by the Swedish Research Council.

\end{document}